\documentclass[prb,twocolumn,showpacs,superscriptaddress,floatfix]{revtex4-1}

\usepackage{epsfig}
\usepackage{amsmath}
\usepackage{amssymb}
\usepackage{amsfonts}
\usepackage{mathptmx}
\usepackage{eucal}
\usepackage{bm}
\usepackage{color}
\usepackage[colorlinks,linkcolor=blue,citecolor=blue]{hyperref}

\usepackage{epstopdf}

\usepackage{graphicx}

\usepackage{natbib}

\begin{document}
\title{Phase-dependent heat transport through magnetic Josephson tunnel junctions }
\author{F. S. Bergeret}
\email{sebastian\_bergeret@ehu.es}
\affiliation{Centro de F\'{i}sica de Materiales (CFM-MPC), Centro
Mixto CSIC-UPV/EHU, Manuel de Lardizabal 4, E-20018 San
Sebasti\'{a}n, Spain}
\affiliation{Donostia International Physics Center (DIPC), Manuel
de Lardizabal 5, E-20018 San Sebasti\'{a}n, Spain}
\affiliation{Institut f\"ur Physik, Carl von Ossietzky Universit\"at, D-26111 Oldenburg, Germany}
\author{F. Giazotto}
\email{giazotto@sns.it}
\affiliation{NEST, Instituto Nanoscienze-CNR and Scuola Normale Superiore, I-56127 Pisa, Italy}
\begin{abstract}
We present an exhaustive study of the coherent heat transport through superconductor-ferromagnet(S-F) Josephson junctions including  a spin-filter (I$_{sf}$) tunneling barrier. 
By using the quasiclassical Keldysh Green's function technique we derive a general expression for the heat current flowing through a S/F/I$_{sf}$/F/S junction and analyze  the dependence of the  thermal conductance on the spin-filter efficiency,  the phase difference between the superconductors and the magnetization direction of the ferromagnetic layers. In the case of non-collinear magnetizations we  show explicitly the contributions to the heat current stemming from the singlet and triplet components of the superconducting condensate. We also demonstrate that  the magnetothermal resistance ratio of a S/F/I$_{sf}$/F/S heat valve can be increased by the spin-filter effect under suitable conditions.
\end{abstract}
\maketitle
\section{Introduction}
Two fields have been  attracting  increasing  attention among several research  groups in  the recent years:  Spintronics with superconductors \cite{Mersevey1994,Zutic2004,BVErmp,EschrigPhysToday} and coherent caloritronics \cite{GiazottoNature,simmonds2012,martinezrectifier2013,golubev,Meschke2006,Vinokur2003,Eom1998,Chandrasekhar2009,Ryazanov1982,Panaitov1984,virtanen2007,rabani1,rabani2,heikkilagiazotto,Ojanen,Ruokola,Pascal}.  Both fields exploit  phase-dependent phenomena which are key characteristics of    superconducting circuits. 
On the one hand, superconducting spintronics  is emerging as a possible technology from  the discovery  of spin-polarized supercurrents\cite{BVErmp} in superconductor-ferromagnet (S/F) hybrid nanostructures.
Such supercurrents are due to existence of triplet superconducting correlations created by magnetic inhomogeinities \cite{BVE01}. 
Once generated, triplet correlations can penetrate over long distances into ferromagnets  as observed in experiments on S/F/S Josephson junctions\cite{Kaizer06, BlamireScience,Birge, Aarts}. These experiments suggest the possibility of using S/F hybrids in spintronic circuits with the aim of lowering the dissipation\cite{EschrigPhysToday}.  
On the other hand,  the study of heat transport in nanoscale devices, {\it i.e.} caloritronics, also attracts the attention of researchers working on nanodevies \cite{Giazotto2006,Dubi2011,muhonen} containing for example normal metal, ferromagnets\cite{bauer2012,Frefrigerator} and superconductors \cite{heattransistor,ser}.  
Of particular interest is the recent experimental control of  the heat current flowing through a Josephson junction  by tuning the macroscopic phase-difference between two superconducting reservoirs\cite{GiazottoNature,giazotto2012,Martinez2013}, as predicted in several theoretical works \cite{Maki1965,Guttman97,Guttman98,Zhao2003,Zhao2004}.  

\begin{figure}[t]
\includegraphics[width=\columnwidth]{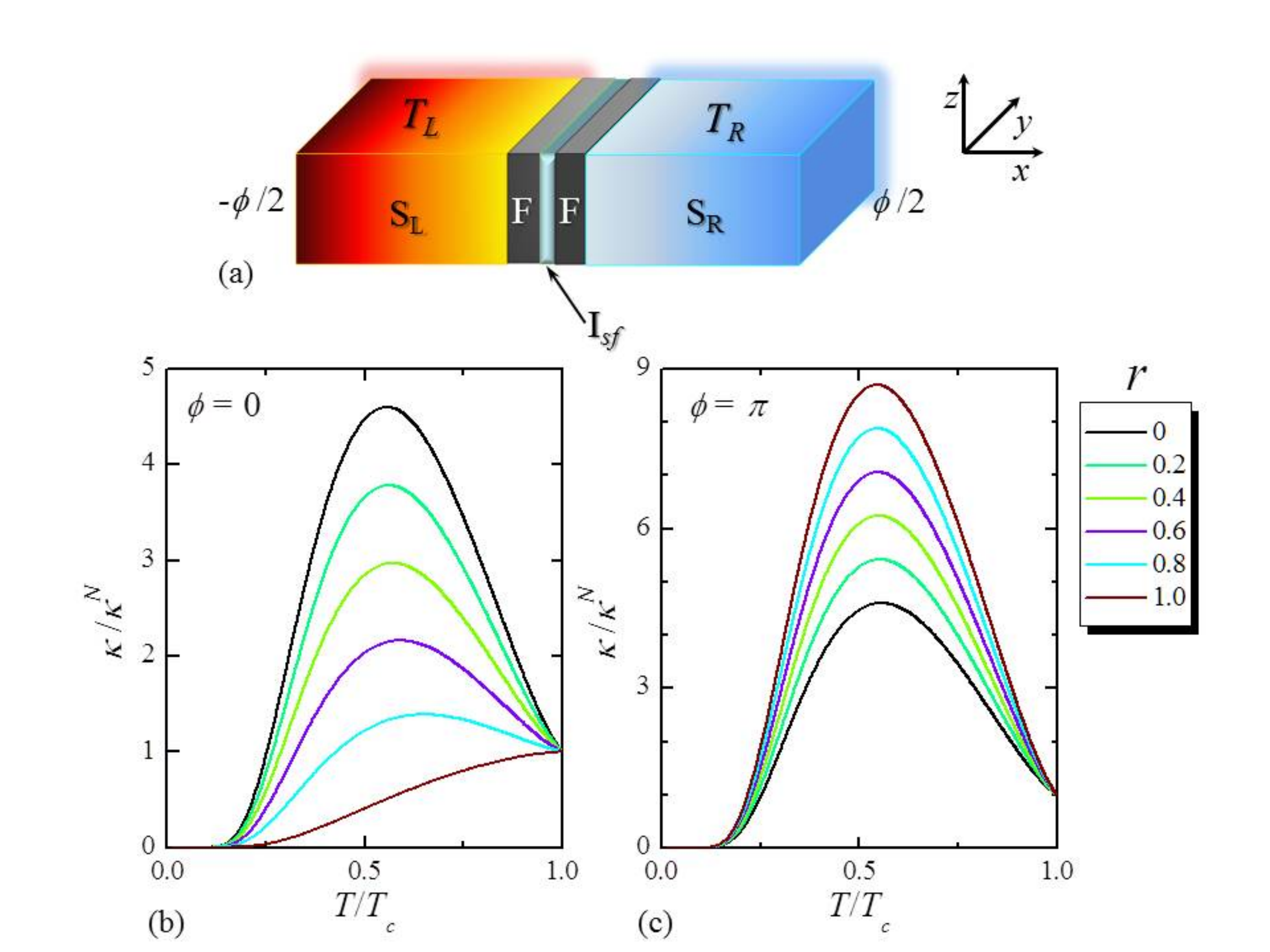}\vspace{-3mm}
\caption{(Color online) (a) Sketch of the generic  SFI$_{sf}$FS Josephson tunnel junction. It  consists of two identical SF layers coupled by a spin-filter barrier I$_{sf}$.  $T_L$ and $T_R$ denotes the temperatures of the superconductors whereas $\phi$ is the macroscopic quantum phase difference over the junction. 
(b) Thermal conductance $\kappa$ vs temperature $T$ calculated for different values of $r$ at $\phi=0$ in the absence of Zeeman splitting in the superconducting electrodes.
(c) The same quantity as in panel (b) calculated for $\phi=\pi$. $\kappa^N$ denotes the thermal conductance in the normal state while $T_c$ is the superconducting critical temperature.}
\label{fig1}
\end{figure}
The interplay between superconductivity and ferromagnetism in the context of heat transport has been recently used to describe
a phase-tunable heat-valve in a recent theoretical work. \cite{Giazotto2013} . The valve is a F/S/I/S/F Josephson junction ($I$ denotes a non-magnetic tunneling barrier) and its  operating principle   is based on  both phase-coherence and  spin-dependent transport. 
Moreover, it is well known that in junctions containing S/F elements
 both  singlet and triplet pair correlations are  generated and contribute  to the Josephson (charge) current and, as we will prove below, to  the phase-dependent part of the heat current. 
If  a spin-filter with a large efficiency  is used as a tunneling barrier, the singlet contribution to the Josephson current is suppressed and a highly spin-polarized supercurrent can be achieved in a S/F/I$_{sf}$/F/S junction provided that  the magnetizations are  non-collinear\cite{Bergeret2012a}(I$_{sf}$ denotes the spin-filter tunneling barrier). As we shall show in the present work, this also applies for the phase-coherent part of the heat current flowing through  a S/F/I$_{sf}$/F/S junction.
 
The spin-filter effect has been intensively studied in europium	chalcogenides	tunneling	barriers.\cite{Moodera90,Moodera08,Moodera2012} This type of barriers possess very large spin-filter efficiencies (typically larger tan  95\%) and, therefore, they are ideal candidates for the creation of spin-polarized currents. 
In tunnel junctions made of superconducting electrodes and spin-filter  barriers, measurements of the tunneling conductance have revealed  that the interaction between conducting electrons in the leads and the localized magnetic moments of the barrier lead to a Zeeman-splitting in the density of states of the superconducting electrodes\cite{giazottotaddei,Moodera90}, as theoretical expected. \cite{Sauls88a} 
  An experiment performed on NbN/GdN/NbN junctions  has shown that the temperature dependence of the Josephson current flowing   through a GdN    barrier (with a spin-filter efficiency of  $\sim75\%$) clearly deviates from that expected in conventional S/I/S junctions\cite{Blamire}, thus suggesting an interplay between magnetism of the barrier and superconducting condensate of the electrodes, as described recently in a theoretical work.\cite{Bergeret2012a}

In the present work we  combine ideas from S/F hybrid structures   and caloritronics studies in order to analyze  the phase-dependent heat transport through such structures.  
We extend  the model proposed in Ref. \cite{Bergeret2012b} for the heat transport   through S/I$_{sf}$/S and F/S/I$_{sf}$/S/F junctions  and derive compact expressions for the thermal conductance. 
With the help of our model we are able to study in detail the  dependence of the heat conductance  on  the spin-filter efficiency, the superconducting phase  and the relative angles between the magnetization of the ferromagnetic layers. 
In analogy to the charge supercurrent we shall demonstrate that the phase-dependent part of the heat current consists of two contributions stemming from singlet and triplet pair correlations, respectively. Moreover, as for the charge transport studied in Ref. \cite{Bergeret2012a}, the spin-filter effect suppresses the singlet contribution to thermal transport leading to spin-polarized heat currents.  
Finally, we  show how the spin-filter barriers can be used for the enhancement of the magnetothermal resistance of Josephson heat valves as those recently proposed  in Ref. \cite{Giazotto2013}.

The paper is organized as follows: In Sec. \ref{model} we  derive a general  expression describing   the heat current  flowing
through a generic  spin-filter junction.  With the help of this expression, in section \ref{results} we  first  analyze 
the heat conductance through a  a S/I$_{sf}$/S junction  as the one used in the  experiments. \cite{Moodera90,Blamire} 
We  demonstrate that while for a zero-phase difference between the superconductors the thermal conductance increases by increasing the spin-filter efficiency, 
 the opposite regime is achieved if the phase difference $\phi$ equals to $\pi$. 
This behavior holds in the presence of  a Zeeman splitting  in the superconductors and also if we neglect this field. 
We also show  that for a large spin-filter efficiency of the barrier the  maximum value of the thermal conductance depends non-monotonically on the amplitude of the Zeeman splitting. 
 In section \ref{results}B  we  consider   a triplet Josephson junction consisting of a F/S/I$_{sf}$/S/F structure, for which the magnetization direction of the outer F layers can point in arbitrary direction with respect to the spin quantization axis determined by the magnetization of the I$_{sf}$ barrier.  
We explicitly  show the contributions of the singlet and triplet part of the condensante to the heat conductance. 
 In  section \ref{results}C we   discuss the  ferromagnetic Josephson thermal valve and show that the magnetothermal resistance ratio in the structure can reach  values as large as $10^6-10^8$ at low temperature depending on the macroscopic phase and on the spin-filter efficiency of the barrier. Finally, we summarize our results in Sec. \ref{summary} .

\section{The Model}
\label{model}

We consider the generic Josephson junction sketched in Fig. \ref{fig1}(a). It consists of two S/F electrodes tunnel-coupled by a spin-filter barrier I$_{sf}$. The thin F layers may  model the effective exchange field induced in the S electrodes due to the presence of the magnetic barrier.\cite{Bergeret2012a} 
This model is accurate if one assume that the F and S layers are in good electric contact and their thicknesses are small enough. \cite{Bergeret2001}
The junction is phase- and temperature-biased. The phase difference between the left (L) and right (R)  electrode is denoted by $\phi$, while their temperatures are kept constant,   at $T_L$ and $T_R$, respectively. 
In order to describe the electronic transport in the junction we introduce the quasiclassical Green's functions (GFs) in the L and R electrodes which are 8$\times$8 matrices in the Nambu-spin-Keldysh space:
 \begin{equation}
{\bf G}_{R(L)}=\left( \begin{array}{cc}
\check G^R_{R(L)} & \check G^K_{R(L)} \\
0 & \check G^A_{R(L)}
 \end{array} \right), 
\end{equation}
where $\check G^{R,A,K}$ are the retarded, advanced and Keldysh components, respectively, which are 8$\times$8 matrices in the Nambu-spin space. 

The expression for the charge current $I_q$ taking into account the spin-filter effect  was derived in Refs.\cite{Bergeret2012a,Bergeret2012b} and reads
\begin{equation}
I_q=[16eR_{N}(\mathcal{T}^{2}+\mathcal{U}^{2})]^{-1}\int d\epsilon \mathrm{%
Tr}\left\{ \hat{\tau}_{3}\left[ {\check{\Gamma%
}}\,\check{G}_{R}(\epsilon ){\check{\Gamma}}^{\dagger },\check{G}_{L}(\epsilon )\right] ^{K}\right\},
\label{Icharge}
\end{equation}
 where $\mathcal{T}$ and $\mathcal{U}$ are the tunneling spin-independent and spin-dependent  matrix elements (for simplicity we neglect their momentum dependence),   
 $\check \Gamma=\mathcal{T}+\mathcal{U}\tau_3\otimes\sigma_3$, $R_{N}=[4\pi e^{2}N_{L}(0)N_{R}(0)(\mathcal{T}^{2}+%
\mathcal{U}^{2})]^{-1}$ is the junction resistance in the normal state,  $N_{R(L)}$ are the density of the states at the Fermi level in the left or right electrode, respectively, and $e$ is the electron charge. 
In analogy and following the derivation carried out in Ref. \cite{Bergeret2012b}one can demonstrate that  the heat current $\dot Q$ is given by 
\begin{equation}
\dot Q=[16e^2R_{N}(\mathcal{T}^{2}+\mathcal{U}^{2})]^{-1}\int d\epsilon \epsilon \mathrm{%
Tr}\left\{ \left[ {\check{\Gamma%
}}\,\check{G}_{R}(\epsilon ){\check{\Gamma}}^{\dagger },\check{G}_{L}(\epsilon )\right] ^{K}\right\}.
\label{Iheat}
\end{equation}
 The GF's in Eqs. (\ref{Icharge}-\ref{Iheat})  have the general structure
\begin{eqnarray} 
\check G^{R(A)}_{R(L)}&=&\hat g^{R(A)}\tau_3+\hat f^{R(A)}(\cos(\phi/2)i\tau_1 \pm\sin(\phi/2)i\tau_2)\label{genGF1}\\
\check G^{K}_{R(L)}&=&(\check G^{R}-\check G^{A})\tanh(\frac{\epsilon}{2T_{R(L)}})\label{genGF2},\; 
\end{eqnarray}
where  $\tau_{1,2,3}$ are the Pauli matrices in Nambu space, $\hat g^{R(A)}$ is the normal and $\hat f^{R(A)}$ the anomalous component of the retarded (advanced) GFs. 
The latter are 2$\times$2 matrices in the spin-space and  are determined by solving the quasiclassical equations in the F/S electrodes. 
Thus, both $I_q$ and $\dot Q$ are given by Eqs. (\ref{Icharge},\ref{Iheat}) after  substituting the  values of the GFs at the interface.\cite{Volkov2008}   
For simplicity we  assume that the thickness of the S and F layers ($t_S$,$t_F$) is  smaller than the characteristic length over which the GFs vary.  
In such a case one can average the quasiclassical equations over the thickness of the F/S bilayer that is now described by an effective exchange field  ($h$) and superconducting order parameter ($\Delta$) defined  by\cite{Bergeret2001} $h/h_0=N_F(0) t_F(N_S(0)t_S+N_F(0)t_F)^{-1}$ and $\Delta/\Delta_0=N_S(0)t_S(N_S(0)t_S+N_F(0) t_F)^{-1}$, respectively. 
In the expressions above, $h_0$  is the bare exchange field existing in each ferromagnetic layer, $\Delta_0$ the bulk superconducting energy gap, and $N_{F,S}(0)$ is the density of states at the Fermi level in the F or S layer, respectively. 
 The normal and anomalous functions in Eqs. (\ref{genGF1},\ref{genGF2}) are given by\cite{BergeretPRB2001} (we skip the upper indices R and A) :
\begin{eqnarray}
\hat g&=&\frac{g_++g_-}{2}+\frac{g_+-g_-}{2}\sigma_3\\
\hat f&=&{f_s}+f_t \sigma_3,
\end{eqnarray}
where $f_s=(f_++f_-)/2$ is the singlet, and  $f_t=(f_+-f_-)/2$ is the triplet (with vanishing total spin projection) components of the condensate, and 
\begin{eqnarray}
 g^{R}_\pm &=&\frac{(\epsilon\pm h)}{\sqrt{(\epsilon\pm h+i\eta)^2-\Delta^2}}\label{gpm}\\
 f^R_\pm &=&\frac{\Delta}{\sqrt{(\epsilon\pm h+i\eta)^2-\Delta^2}}\label{fpm}.
\end{eqnarray}
 Same expressions hold for the advanced GFs if we substitute $i\eta$ by $-i\eta$. The latter parameter describes the inelastic scattering rate within the relaxation time approximation\cite{relaxation} and it is set $\eta=10^{-5}\Delta_0$ throughout the article. The density of the states of the electrodes is given by the real part of $g_+^R+g_-^R$.
 Notice that the order parameter $\Delta$ in Eqs. (\ref{gpm},\ref{fpm}) has to be calculated self-consistently 
from the gap equation 
$\text{ln}(\Delta_0/\Delta)=\int_0^{\hbar\omega _D}d\varepsilon (\varepsilon ^2+\Delta^2)^{-1/2}[f_+(\varepsilon)+f_-(\varepsilon)]$, where $f_{\pm}(\varepsilon)=\left\{1+\text{exp}[\frac{1}{T}(\sqrt{\varepsilon ^2+\Delta^2}\mp h)]\right\}^{-1}$ and $\omega_D$ is the Debye frequency of the superconductor.  Eqs. (\ref{Iheat}-\ref{fpm}) are used in the next sections in order to analyze the heat transport through a variety of tunneling junctions based on the prototypical example of Fig. \ref{fig1}(a). 

\section{Results}
\label{results}
We now use the above derived equations to determine the heat transport through Josephson junctions with spin filters.  While the charge current (quasiparticle and Josephson components) in such structures has been analyzed both experimentally ( in  Al/EuS/Al  \cite{Moodera90}, and NbN/GdN/NbN\cite{Blamire} junctions) and theoretically discussed \cite{Bergeret2012a,Bergeret2012b}, heat transport in S/I$_{sf}$/S has not be studied so far.   
In what follows we present the results for the thermal conductance, $\kappa=\dot Q/\delta T$, in different structures.  
$\kappa$  can be obtained from  Eq. (\ref{Iheat}), and in the case of identical electrodes   is given by 
\begin{equation}
\label{kappa}
\kappa =\frac{1}{2e^2R_{N}}\sum_{\alpha=\pm}\int d\epsilon \epsilon.\left(\frac{\partial F}{\partial T}\right) \left\{ N_\alpha^2- rM_\alpha^2\cos\phi\right\},
\end{equation}
 where $\delta T=T_L-T_R$,  $(\partial F/\partial T)=-\epsilon/[2T^2\cosh^2(\epsilon/2T)]$, $N_\alpha=(g^R_\alpha-g^A_\alpha)/2$, $M_\alpha=(f^R_\alpha-f^A_\alpha)/2$, $r=\frac{\mathcal{T}^2-\mathcal{U}^2}{\mathcal{T}^2+\mathcal{U}^2}$, and we have assumed that  $\delta T\ll T=(T_R+T_L)/2$.  The parameter $r$ is a measure for the spin-filter efficiency ${\mathcal{P}}=\sqrt{1-r^2}$ of the barrier: it is equal to 0 for a 100\% spin-filter efficiency and $r=1$ for a non-magnetic barrier.  The second term in the r.h.s of Eq. (\ref{kappa}) is the phase-dependent anomalous  term, which was obtained for the first time by Maki and Griffin.\cite{Maki1965}.  According to Eq. (\ref{kappa}) the phase-coherent contribution to $\kappa$ is suppressed  by increasing the spin-filter efficiency, {\it i.e.}, by decreasing $r$.  The fact that an  increasing spin-filter efficiency blocks gradually  the phase-dependent contribution to the heat current, demonstrates that the latter is due to electron  pairs with different spin orientation.   As we shall show below, if we allow for triplet pairs with finite total spin projection, the phase-dependent contribution to $\kappa$ does not vanish even if $\mathcal{P}=1$. 
\begin{figure}[tb]
\includegraphics[width=\columnwidth]{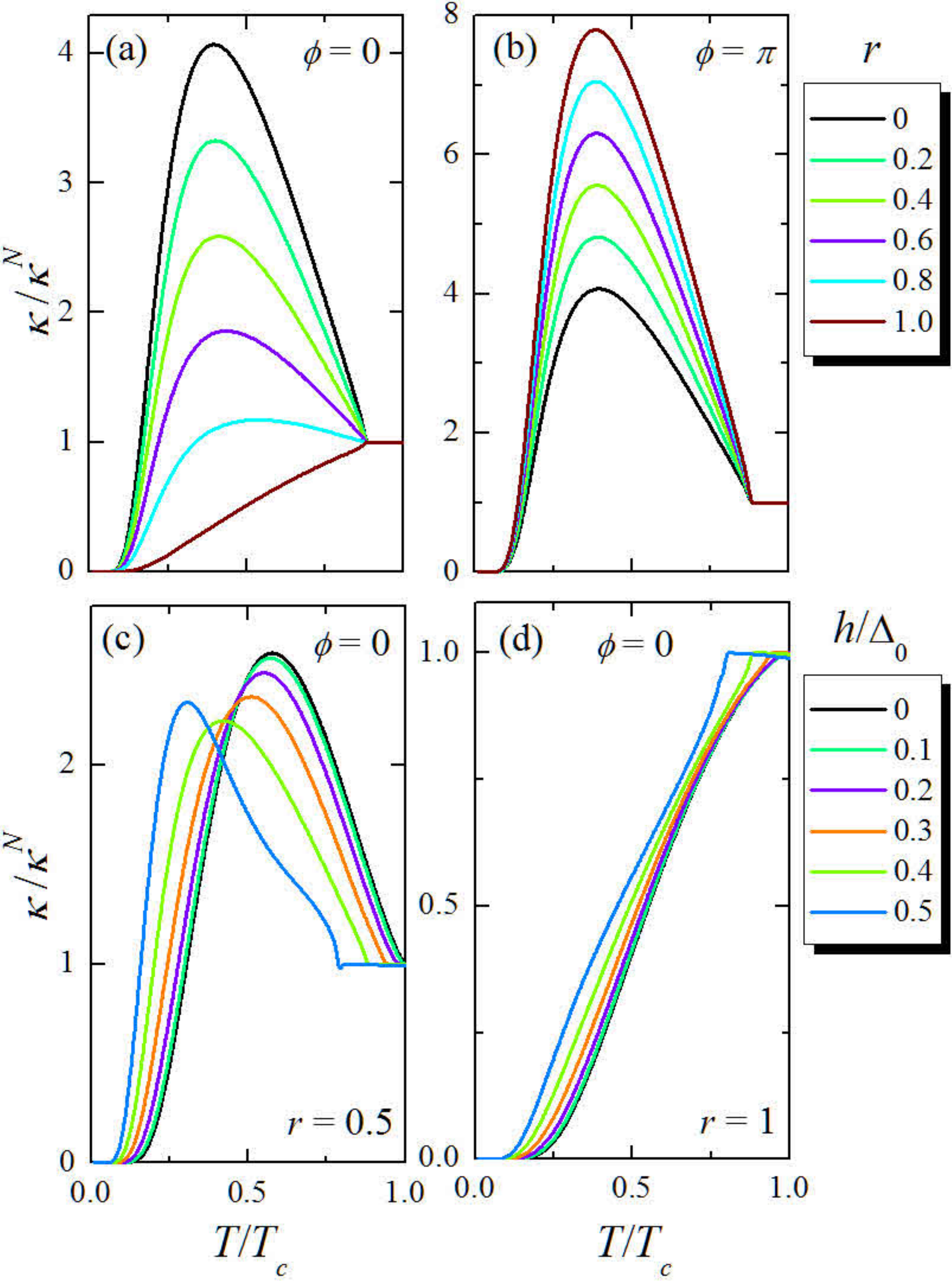}\vspace{-3mm}
\caption{(Color online) (a) Thermal conductance  vs $T$ calculated for several values of $r$ at $\phi=0$. 
(b) The same quantity as in panel (a) calculated for $\phi=\pi$. In panels (a) and (b) we set $h=0.4\Delta_0$ where $\Delta_0$ is the zero-temperature, zero-exchange field superconducting order parameter.
(c) Thermal conductance  vs $T$ calculated for a few values of $h$ at $\phi=0$ and for finite spin-filter efficiency ($r=0.5$). 
(d) The same quantity as in panel (c) calculated in the absence of a spin-filter barrier ($r=1$). }
\label{fig2}
\end{figure}

\subsection{The S/I$_{sf}$/S Junction}

We start our analysis by considering a simple  S/I$_{sf}$/S junction. We first assume 
that there is no exchange field induced in the S electrodes. 
This occurs when the coupling between the conducting electrons in  the superconducting leads and the magnetic moments localized at the barrier can be neglected, for instance, due the presence of a non-magnetic oxide  between the   I$_{sf}$  and S layers\cite{Moodera90}.  
In  such a case,  one can set in Eq. (\ref{kappa}) $N_+=N_-$ and $M_+=M_-$.  
Figures \ref{fig1}(b) and (c)  show  the temperature dependence of $\kappa$  for two values of $\phi$ and different spin-filter efficiencies. 
Throughout the paper the thermal conductance is shown normalized to that in the normal state, $\kappa^N=\mathcal{L}_0T/R_N$, where $\mathcal{L}_0=\pi^2k_B^2/3e^2$ is the Lorenz number and $k_B$ is the Boltzmann constant.
If $\phi=0$  the contribution to $\kappa$ from the phase-dependent part is negative, and therefore by decreasing $r$, ({\it i.e.}, by increasing the efficiency of the spin-filter) the thermal conductance increases [see Fig. \ref{fig1}(b)]. 
On the contrary, for $\phi=\pi$ the anomalous contribution to $\kappa$  is positive, and the thermal conductance decreases with $r$. With the exception of $r=1$ and $\phi=0$ case, $\kappa$ always shows a maximum at a certain finite temperature  ($T\approx0.55 T_C$). 

If we now assume  a good I$_{sf}$/S contact and thin S layers the density of states of the latter shows a  Zeeman splitting which acts as an effective exchange field $h$ inside the superconductor in accordance with Eq. (\ref{gpm}).    This is  induced by  the magnetic proximity effect of the $I_{sf}$ barrier\cite{Moodera90,Sauls88a}. 
We note that our model can  also describe S/F/ I$_{sf}$ /F/S structures with two  thin ferromagnetic films [see Fig. \ref{fig1}(a)].  
In Figs. \ref{fig2}(a) and \ref{fig2}(b) we have chosen $h=0.4\Delta_0$ and calculated the temperature dependence of $\kappa$ for $\phi=0$ and  $\phi=\pi$, respectively.  Due to the presence of the exchange field the superconducting  critical temperature of the SF electrodes is reduced by a factor $\sim 0.875$ with respect to the bulk $T_c$.  
The black curves in Figs. \ref{fig2}(a) and \ref{fig2}(b) correspond to a  perfect spin-filter with $\mathcal{P}=1$ ($r=0$). According to Eq. (\ref{kappa}), in this case,  the only contribution to $\kappa$ comes from the quasiparticle channel.
As in the zero exchange field case, if $r\neq 0$ the corrections to $\kappa$  from the phase-dependent anomalous term in  Eq. (\ref{kappa})  are negative for $\phi=0$ and positive for $\phi=\pi$. This explains why for $\phi=0$  the amplitude of the thermal conductance decreases by increasing $r$ [see Fig. \ref{fig2}(a)], whereas for $\phi=\pi$ the thermal conductance increases with $r$ [see Fig. \ref{fig2}(b)]. 

In panels  (c) and (d) of Fig. \ref{fig2} we compare  the $\kappa(T)$ dependence in the presence and in the absence, respectively,  of a spin-filter barrier. 
Here we set a zero phase difference, $\phi=0$.  If the tunneling barrier is non-magnetic, $r=1$,  the transition to the superconducting state leads to a decrease of the thermal conductance as shown in Fig. \ref{fig2}(d).  
Notably, in this case for any temperature $\kappa$ increases monotonically by enhancing the amplitude of the effective exchange field $h$. By contrast, if the tunneling barrier has a finite spin-filter efficiency ($r=0.5$ which corresponds to $\mathcal{P}\approx0.88$), below the superconducting transition temperature, $T\lesssim T_c$, the thermal conductance increases by decreasing the exchange field. 
By further decreasing the temperature,  $\kappa$  shows a maximum,  and then decays to zero [see Fig. \ref{fig2}(c)]. The maximum value of $\kappa$ ($\kappa_{max}$) depends non-monotonically on $h$:
For small enough  values of $h$,  $\kappa_{max}$ decreases  by increasing $h$, however for $0.4\Delta_0<h<0.5\Delta_0$ it turns out to increase. 

From Eq. (\ref{kappa}) it  clearly appears that for a spin-filter with 100\% efficiency ($r=0$),   the anomalous contribution to $\kappa$ vanishes [i.e., the last term in Eq. (\ref{kappa} is zero]  and therefore  the heat transport will not depend on the phase difference $\phi$. 
\begin{figure}[t]
\includegraphics[width=\columnwidth]{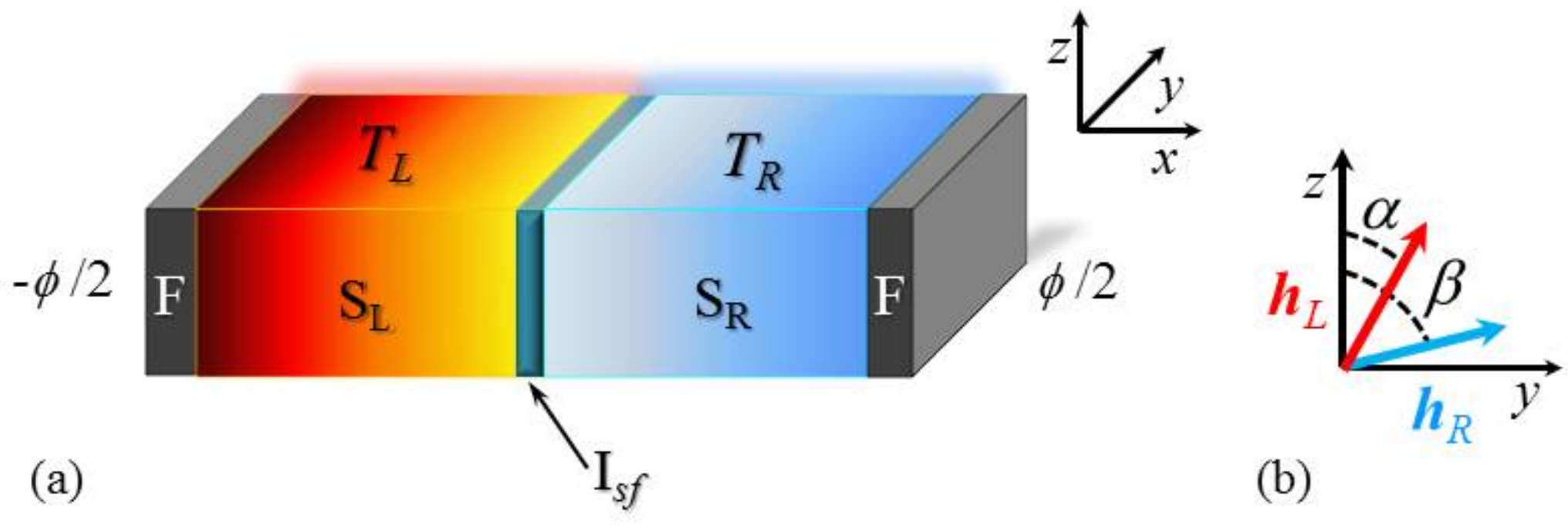}\vspace{-3mm}
\caption{(Color online) (a) The triplet  FSI$_{sf}$SF Josephson tunnel junction discussed in the text. The two FS bilayer are tunnel-coupled by a spin-filter barrier I$_{sf}$.
(b) The exchange fields ($h_{L,R}$) in the ferromagnetic layers are confined to the $y-z$ plane, and are misaligned by an angle $\alpha$ and $\beta$, respectively, with respect to the  $z$-axis. The latter is defined by  the magnetization direction of the I$_{sf}$ barrier.}
\label{fig3}
\end{figure}

\subsection{Triplet Josephson junctions with spin-filter}

In order to detect the spin triplet supercurrents, long-range Josephson effect  has been measured in a variety of multilayered ferromagnetic structures\cite{BlamireScience,Birge,Robinson2010} with inhomogeneous magnetic configurations. According to the theoretical prediction\cite{BVErmp} , such inhomogeneity induces the  triplet pair correlations with equal spin-projection in the ferromagnetic bridge. Here we aim to understand the heat transport through  S/F hybrid structures containing tunneling barriers. For that sake we consider the  structure shown
 in Fig. \ref{fig3}(a). It consists of two FS bilayers tunnel-coupled by a spin-filter barrier.  
 We set the \emph{z}-axis (spin quantization axis) parallel to the magnetization of the I$_{sf}$ layer, and  define the angles, $\alpha$ and $\beta$, which describe the direction of magnetization of the left and right ferromagnets, respectively [see Fig. \ref{fig3}(b)]. For a good contact between the S and F layers and small enough thicknesses this structure is equivalent to the one shown in Fig. \ref{fig1}(a). 

\begin{figure}[t]
\includegraphics[width=\columnwidth]{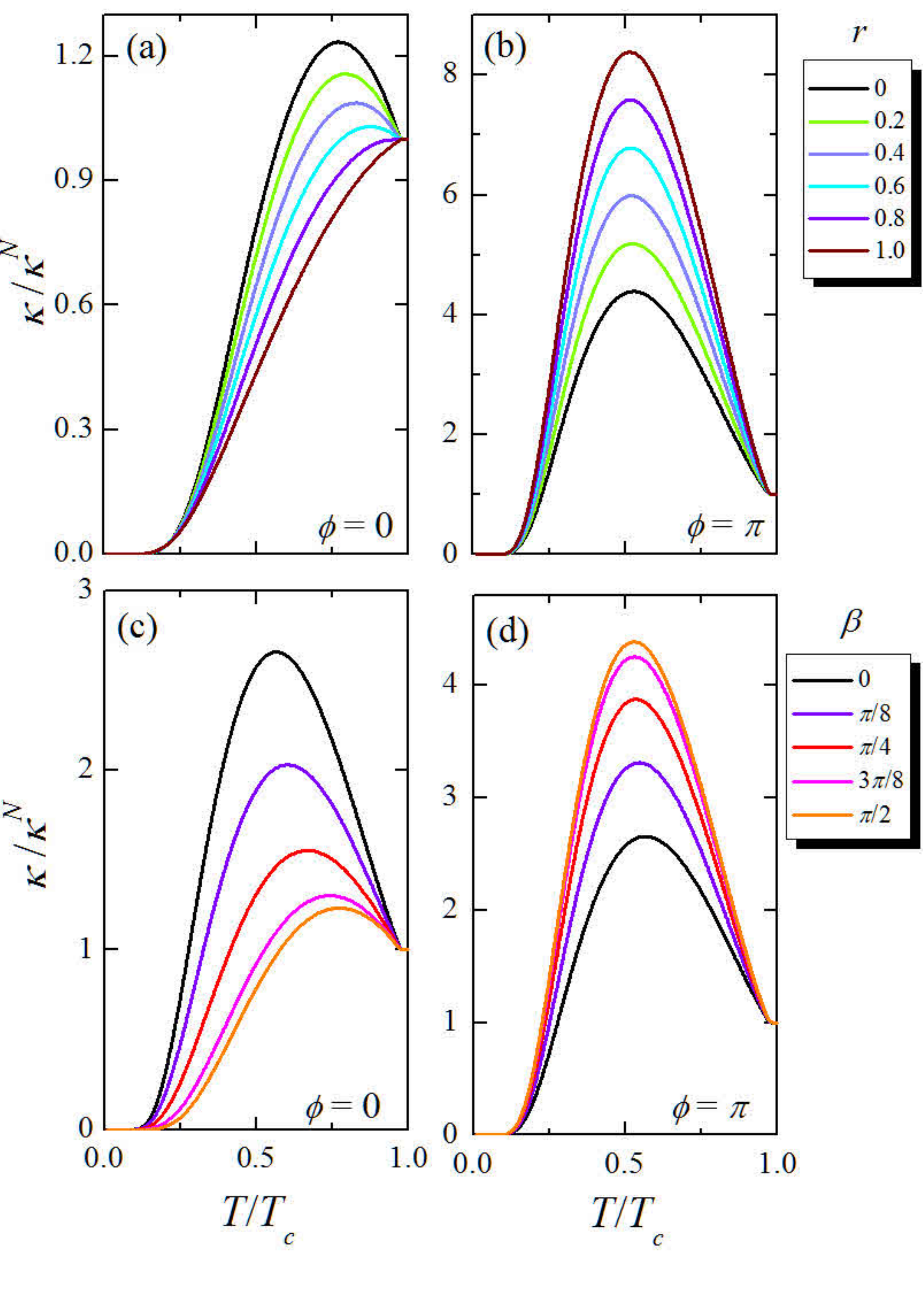}\vspace{-3mm}
\caption{(Color online) (a) Thermal conductance vs $T$ calculated for several values of $r$ at $\phi=0$.
(b) The same quantity as in panel (a) calculated at $\phi=\pi$. In panels (a) and (b) we set $\alpha=\beta=\pi/2$.
(c) Thermal conductance vs $T$ calculated for several values of $\beta$ at $\phi=0$.
(d) The same quantity as in panel (c) calculated at $\phi=\pi$. In panels (c) and (d) we set $r=0$ and $\alpha=\pi/2$. In  all the calculations of the figure we assumed $h=0.2\Delta_0$.}
\label{fig4}
\end{figure}
The generalized expression for the thermal conductance in this case can be derived from Eq. (\ref{Iheat}) with the help of the  technique used in Refs.\cite{Bergeret2012a, Bergeret2012b}.  We obtain $\kappa=\kappa_{qp}+\kappa_{\phi}$, where $\kappa_{qp}$ is the contribution from the quasiparticles to thermal transport   given by
\begin{eqnarray}
\label{kappaqp}
&\kappa_{qp}& =\frac{1}{4e^2R_{N}}\int d\epsilon \epsilon.\left(\frac{\partial F}{\partial T}\right) \left\{ \left[N_++N_-\right]^2+\right. \\
&+&   \left .  \left[N_+-N_-\right]^2\cos\alpha\cos\beta   + r\sin\alpha\sin\beta[N_+-N_-]^2\right\},\nonumber
\end{eqnarray}
and $\kappa_{\phi}$ is  the anomalous phase-dependent contribution  that can be written in terms of the singlet ($f_s$) and triplet ($f_t$) component of the condensate:
\begin{eqnarray}
\label{kappaph}
\kappa_{\phi} &=&-\frac{\cos\phi}{4e^2R_{N}}\int d\epsilon \epsilon.\left(\frac{\partial F}{\partial T}\right) \left[  rM_t^2\cos\alpha\cos\beta+\right.\nonumber\\
&+&\left. rM_s^2+M_t^2\sin\alpha\sin\beta\right],
\end{eqnarray}
where $M_t\equiv M_+-M_-=f^R_t-f^A_t$ and  $M_s\equiv M_++M_-=f^R_s-f^A_s$.  Notice that even in the case of a perfect spin-filter efficiency ($r=0$) there  is a phase-dependent contribution to $\kappa$ provided that the magnetization of the F layers are non-collinear with the one of the barrier (i.e., $\alpha,\beta\neq0,\pi$).  
In such a case, the measured $\kappa(\phi)$ dependence is a direct manifestation of the triplet component of the condensate in analogy to the finite   charge supercurrent flowing through a fully efficient spin-filter, as recently predicted in  Ref. \cite{Bergeret2012a}. 
Again,  the phase dependent contribution $\kappa_\phi$  is proportional to $\cos\phi$ [{\it cf.} Eq.(\ref{kappa})]  and therefore we expect for $\kappa(T)$ a similar behavior as for the S/I$_{sf}$/S structure. This is confirmed in panels (a) and (b) of Fig. \ref{fig4} where we show the temperature dependence of $\kappa$ for the F layers having a  magnetization  parallel to each other but perpendicular to the magnetization of the barrier, {\it i.e.}  $\alpha=\beta=\pi/2$.  
In particular, the thermal conductance can increase considerably with respect to the normal value if $\phi=\pi$.  
In panels (c) and (d) of Fig. \ref{fig4} we show the temperature dependence of $\kappa$ for different angles $\beta$ by setting $\alpha=\pi/2$. In the $\phi=0$ case  maximum values for $\kappa$ are achieved for $\beta=0$, whereas if $\phi=\pi$ the maximum $\kappa$ is observed for $\beta=\pi/2$.

\begin{figure}[t]
\includegraphics[width=\columnwidth]{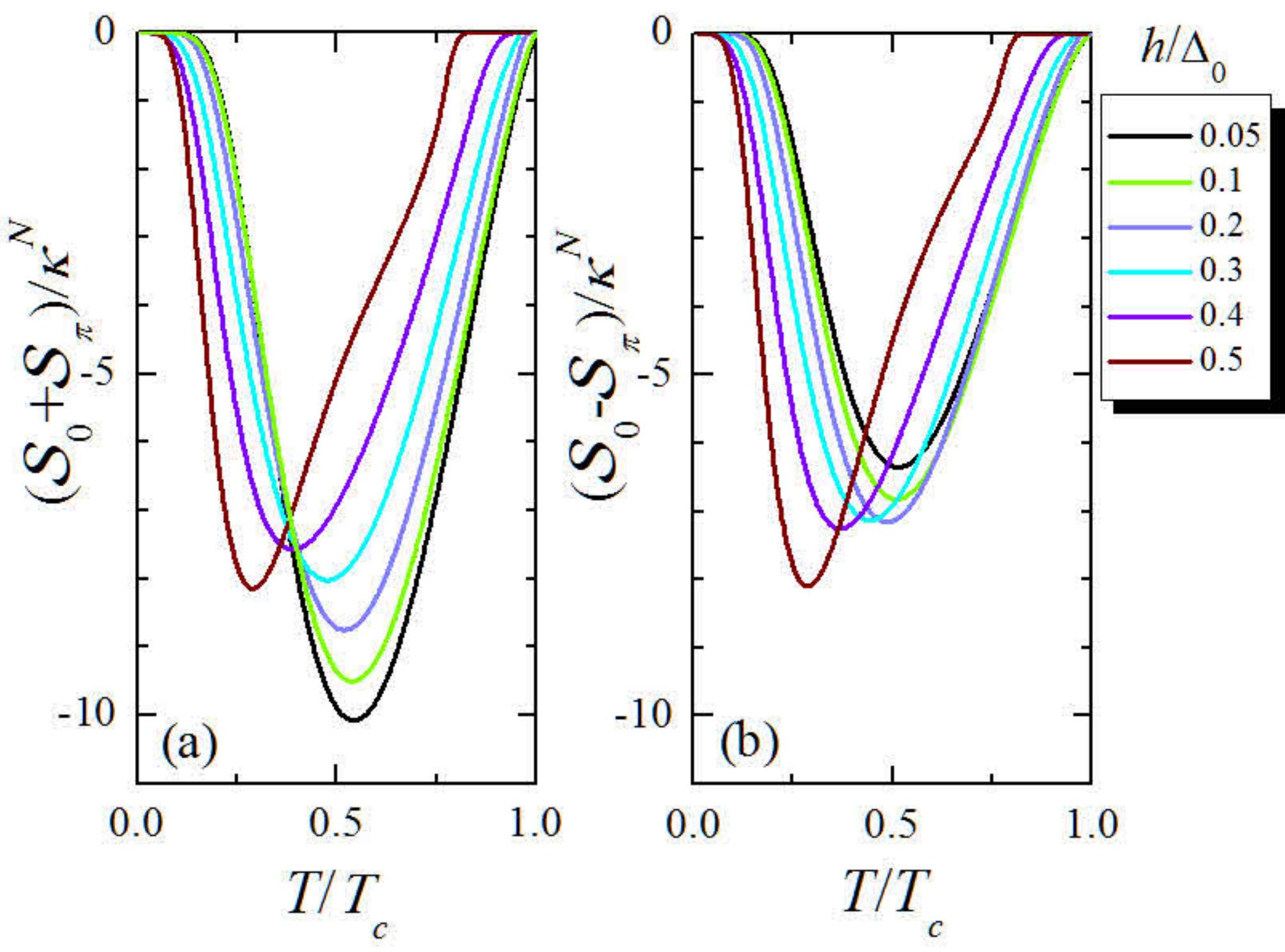}\vspace{-3mm}
\caption{(Color online) (a) Singlet pairs contribution to  thermal conductance  vs $T$ calculated for several values of the exchange field in the ferromagnetic layers.
(b) Triplet pairs contribution to thermal conductance  vs $T$ calculated for the same $h$ values as in panel (a). }
\label{fig5}
\end{figure}
In the case of a perfect spin-filter ($r\rightarrow 0$), one can see  from  Eq. (\ref{kappaph}) that only the  triplet term  $M_t$ contributes to $\kappa$. This term  describes the spin-polarized heat current. 

In principle one can analyze   the contributions from the singlet and  triplet pairs density separately by   consider the junction of Fig. \ref{fig3}(a)  with a non-magnetic tunneling barrier ({\it i.e.}, $r=1$). 
We set  the magnetization of one of the F layer  fixed ({\it e.g.}, $\alpha=0$) and then we switch the other F layer magnetization between a parallel ($\beta=0$) or antiparallel ($\beta=\pi$) configuration.  
If we now  perform a phase-biased experiment \cite{GiazottoNature} and measure $\mathcal{S}_\beta$, i.e., the difference between the heat conductance $\kappa(\phi,\beta)$ for $\phi=0$ and $\phi=\pi$,  
\begin{equation}
\label{S}
\mathcal{S}_\beta=\kappa(0,\beta)-\kappa(\pi,\beta),
  \end{equation}
in the parallel and antiparallel configuration it is clear from Eqs. (\ref{kappaqp}-\ref{kappaph}) that $\mathcal{S}_0+\mathcal{S}_\pi$  represents the contribution from singlet pairs
 \begin{equation}
\mathcal{S}_0+\mathcal{S}_\pi=-\frac{1}{e^2R_N}\int d\epsilon \epsilon.\left(\frac{\partial F}{\partial T}\right)M_s^LM_s^R,
\end{equation}
whereas the difference $\mathcal{S}_0-\mathcal{S}_\pi$ represents the one from triplet pairs
 \begin{equation}
\mathcal{S}_0-\mathcal{S}_\pi=-\frac{1}{e^2R_N}\int d\epsilon \epsilon.\left(\frac{\partial F}{\partial T}\right)M_t^LM_t^R.
\end{equation}
These  two contributions are plotted in Fig. \ref{fig5} as a function of the temperature for different values of the exchange field. 
In particular, the maximum contribution from the singlet component is achieved for the lowest values of the exchange field around $T\sim0.5T_c$, whereas the triplet contribution is maximized by increasing the exchange field value (i.e., in the present case $h=0.5\Delta_0$) around $T\sim0.25T_c$. At large enough exchange fields both contributions tend to be similar. 
We note that  at low temperature the amplitude of the singlet component decreases not monotonically by increasing $h$ whereas that of the triplet contribution turns out to monotonically increase by increasing the exchange field.
\begin{figure}[t]
\includegraphics[width=\columnwidth]{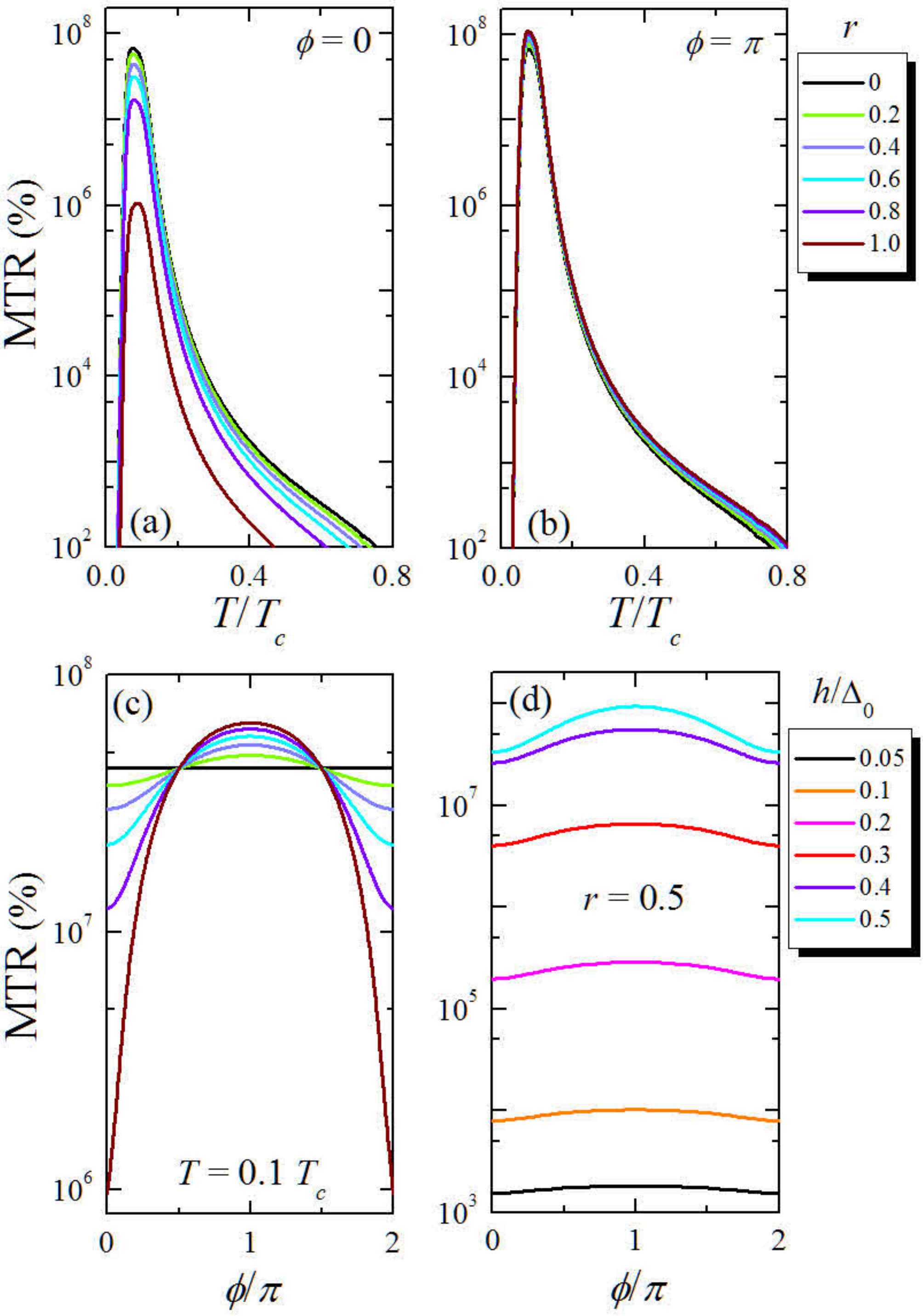}\vspace{-3mm}
\caption{(Color online) (a) Magnetothermal resistance ratio MTR vs temperature $T$ calculated for several values of $r$ at $\phi=0$ and $h=0.4\Delta_0$. 
(b) MTR ratio vs $T$ calculated for the same $r$ values as in panel (a) at $\phi=\pi$ and $h=0.4\Delta_0$.
(c) MTR ratio vs phase calculated for the same $r$ values as in panel (a) at $T=0.1T_c$ and $h=0.4\Delta_0$.
(d) MTR ratio vs phase calculated for a few values of the exchange field $h$ at $T=0.1T_c$ and $r=0.5$.
 \label{fig6}}
\end{figure}
\subsection{The Josephson heat valve}

A similar junction as the one shown in Fig. \ref{fig3}(a) (with a non-magnetic tunneling barrier instead of I$_sf$)  was recently proposed by the authors as a heat valve\cite{Giazotto2013}. It was shown that the electronic contribution to thermal conductance strongly depends on the relative  magnetization angle between the F layers. 
In particular, values for the magnetothermal resistance (MTR) ratio as large as $10^5-10^7$\% has been predicted to occur at low temperature\cite{Giazotto2013}. The MTR ratio can be defined as
\begin{equation}
\textrm{MTR}=\frac{\kappa_P-\kappa_{AP}}{\kappa_{AP}}\; , 
\end{equation}
where P and AP denote the parallel and antiparallel configuration of the magnetization in the F layers, respectively. 
In the context of the present paper a natural question arises: How does a spin-filter barrier affect the MTR ratio? The answer to this question can be found in Fig. \ref{fig6} where we plot the behavior of the MTR as a function of the temperature  and the Josephson phase. 
Figure \ref{fig6}(a) shows that the MTR ratio increases by increasing the spin-filter efficiency for $\phi=0$.  
In such a case,  a 100\% spin-filter efficiency [upper curve in Fig. \ref{fig6}(a)] leads to values of   MTR  which are almost  two orders of magnitude larger than in the absence of a magnetic barrier [$r=1$, lower curve in Fig. \ref{fig6}(a)].  
By contrast, for $\phi=\pi$ the MTR ratio depends only weakly on $r$, and decreases by increasing the spin-filter efficiency.
 The heat valve effect turns out to be maximized for both phases around $T\sim 0.1T_c$.

The  phase dependence of the MTR is plotted   in  the lower panels of Fig. \ref{fig6}.   Figure \ref{fig6}(c) shows this dependence  for  $T=0.1T_c$ and the same $r$ values as in panels (a-b) . 
The MTR ratio is minimized for zero phase difference and reaches its maximum value at  $\phi=\pi$. 
Since we are only considering collinear magnetizations ({\it i.e.}, either parallel or anti-parallel) the phase-dependent contribution to $\kappa$ vanishes if $r=0$ [{\it cf.} Eq. (\ref{kappaph})], and in turn the MTR ratio does not depend on $\phi$,  as shown by the black curve in Fig. \ref{fig6}(c).  
All curves cross at $\phi=\pi/2$, which is the phase value separating the two behaviors: If $0\leq \phi<\pi/2$ the MTR decreases by increasing $r$ while the opposite behavior is achieved for $\pi/2<\phi\leq \pi$.
It is worthwhile mentioning that  in the parallel configuration the Josephson valve heat conductance is maximized. In contrast,  the dc Josephson effect  is maximized by the  anti-parallel configuration.\cite{Bergeret2001}  This means that in the P configuration the ferromagnetic Josephson junction behaves as an almost ideal \emph{electric} insulator  whereas in the AP one  it behaves as an ideal \emph{thermal} insulator\cite{Giazotto2013}.   

Panel \ref{fig6}(d) shows the phase dependence of the MTR ratio calculated for a few different values of $h$ and a moderate spin-filter efficiency $r=0.5$ at $T=0.1T_c$. It clearly appears that the larger the splitting field induced in the S layers, the larger is the  heat valve effect.

\section{Summary}
\label{summary}
In summary, we have presented an exhaustive study of the electronic heat transport in SF/I$_{sf}$/SF  Josephson junctions  with magnetic and non-magnetic I$_{sf}$ tunneling barriers. 
General expressions for the heat current and heat conductance $\kappa$  were derived taking into account the spin-filter efficiency $\cal P$  of the barrier. 
 It has been shown that $\kappa$ strongly depends on $\cal P$. For a given value of the exchange field  two behaviors have been found: 
 In the case of a zero phase difference 
between the SF electrodes an increasing spin-filter efficiency leads to a increase of $\kappa$, whereas the opposite behavior is achieved if $\phi=\pi$. We have also investigated the heat conductance in the case 
that the magnetizations of the F layers and the spin-filter are non-collinear.  
We explicitly computed the contributions to $\kappa$ stemming from   \emph{singlet} and  \emph{triplet} pair correlations. Finally, we have analyzed a heat valve based on a F/S/I$_sf$/S/F Josephson junction, and demonstrated that for $\pi/2<\phi\leq\pi$ the lowering the spin-filter efficiency of the barrier  leads to a sizable  enhancement of the   magnetothermal resistance ratio. 

We finally  discuss here   some potential applications of the analyzed structures.
Ferromagnetic Josephson heat valves  can  be used  whenever a precise tuning and mastering of the temperature is required, for instance, for on-chip heat management as a switchable heat sink. 
Furthermore, such a valve  setup can  be useful as well,   to  tune the operation temperature of radiation sensors. \cite{Giazotto2006,Giazotto2008}
In the context of quantum computation \cite{Nielsen2002}  these elements can also be used to influence the behavior and the dynamics of two-level quantum systems through temperature manipulation. 
Finally, the strong dependence of the Josephson supercurrent on temperature can be exploited for the realization of controllable thermal Josephson junctions of different kinds \cite{Giazotto2006,Giazotto2004,Tirelli2008,Savin2003,giazottoprl2004}.

\acknowledgments
 
The work of F.S.B was supported by the Spanish Ministry of Economy and Competitiveness under Project FIS2011-28851-C02-02. F.S.B thanks Prof. Martin Holthaus and his group for their kind hospitality at the Physics Institute of the Oldenburg University.
F.G. acknowledges the FP7 program No. 228464 "MICROKELVIN", the Italian Ministry of Defense through the PNRM project "Terasuper", and the Marie Curie Initial Training Action (ITN) Q-NET 264034 for partial financial support.

\end{document}